\documentclass[sigplan,screen]{acmart}
\pdfoutput=1

\usepackage{booktabs}   \usepackage{subcaption} \usepackage[utf8]{inputenc}
\usepackage[T1]{fontenc}
\usepackage{microtype}

\usepackage{amsmath,mathtools} \usepackage{hyperref}
\hypersetup{
    breaklinks=true,            }
\usepackage{breakurl}
\usepackage{xspace} \usepackage{graphicx,tikz} \graphicspath{{./fig/}}
\usepackage{cleveref} 

\usepackage{syntax}

\makeatletter
\@namedef{ver@lineno.sty}{9999/12/31}
\@namedef{opt@lineno.sty}{}
\makeatother
\usepackage[draft]{minted}
\usepackage{xintexpr}

\newcommand{\name}{Proverbot9001\xspace}

\newcommand{\coqinline}[1]{\mintinline{coq}|#1|}
\newcommand{\States}{\mathcal{S}}
\newcommand{\Obligations}{\mathcal{O}}
\newcommand{\Hypotheses}{\mathcal{H}}
\newcommand{\Goals}{\mathcal{G}}
\newcommand{\Idents}{\mathcal{I}}
\newcommand{\Props}{\mathcal{Q}}
\newcommand{\Tactics}{\mathcal{T}}
\newcommand{\Reals}{\mathbb{R}}
\newcommand{\Args}{\mathcal{A}}
\newcommand{\Commands}{\mathcal{C}}
\newcommand{\PTac}{P_{\it tac}}
\newcommand{\PArg}{P_{\it arg}}
\newcommand{\PState}{P}
\newcommand{\st}{\sigma}

\newcommand{\tac}{\tau}
\newcommand{\argu}{a}
\newcommand{\Predictors}{\mathcal{R}}
\newcommand{\ScoringFuncs}{\mathcal{D}}
\newcommand{\Combine}{\otimes}
\newcommand{\PH}{138}
\newcommand{\PHPercent}{\number\numexpr((\PH * 100) / 501)\relax\%\xspace}
\newcommand{\CHCP}{142}

\newcommand{\cut}[1]{}
 
\setcopyright{acmlicensed}
\acmPrice{15.00}
\acmDOI{10.1145/3394450.3397466}
\acmYear{2020}
\copyrightyear{2020}
\acmSubmissionID{pldiws20maplmain-p2-p}
\acmISBN{978-1-4503-7996-0/20/06}
\acmConference[MAPL '20]{Proceedings of the 4th ACM SIGPLAN International Workshop on Machine Learning and Programming Languages}{June 15, 2020}{London, UK}
\acmBooktitle{Proceedings of the 4th ACM SIGPLAN International Workshop on Machine Learning and Programming Languages (MAPL '20), June 15, 2020, London, UK}

\begin{document}
\sloppy
\catcode`\_=11

\title{Generating Correctness Proofs with Neural Networks}

\author{Alex Sanchez-Stern}
\affiliation{
  \institution{UC San Diego}            \country{USA}
}
\email{alexss@eng.ucsd.edu}          

\author{Yousef Alhessi}
\affiliation{
  \institution{UC San Diego}            \country{USA}
}
\email{yalhessi@eng.ucsd.edu}          

\author{Lawrence Saul}
\affiliation{
  \institution{UC San Diego}            \country{USA}
}
\email{saul@cs.ucsd.edu}          

\author{Sorin Lerner}
\affiliation{
  \institution{UC San Diego}            \country{USA}
}
\email{lerner@cs.ucsd.edu}

\begin{abstract}
Foundational verification allows programmers to build
  software which has been empirically shown to have high levels of assurance
  in a variety of important domains.
However, the cost of producing foundationally verified software
  remains prohibitively high for most projects,
  as it requires significant manual effort by highly trained experts.
In this paper we present \name{},
  a proof search system using machine learning techniques
  to produce proofs of software correctness
  in interactive theorem provers.
We demonstrate \name{} on the proof obligations
  from a large practical proof project,
  the CompCert verified C compiler,
  and show that it can effectively automate
  what were previously manual proofs,
  automatically producing proofs for \PHPercent of theorem statements
  in our test dataset, when combined with solver-based tooling.
Without any additional solvers,
  we exhibit a proof completion rate that is a 4X improvement
  over prior state-of-the-art machine learning models
  for generating proofs in Coq.

\end{abstract}



\maketitle

\section{Introduction}
\label{sec:introduction}

A promising approach to software verification is \emph{foundational
  verification}. In this approach, programmers use an interactive
theorem prover, such as Coq~\cite{coq} or
Isabelle/HOL~\cite{isabelle}, to state and prove properties about
their programs. \cut{The proofs are performed interactively via the use of
\emph{proof commands}, which programmers invoke to make progress on a
proof. To complete a proof, a programmer must provide guidance to the
proof assistant at each step by picking which proof command to
apply.} Foundational verification has shown increasing promise over the
past two decades; it has been used to prove properties of programs in
a variety of settings, including compilers~\cite{compcert}, operating
systems~\cite{sel4}, database systems~\cite{verified-db}, file
systems~\cite{verified-fs}, distributed systems~\cite{verdi}, and
cryptographic primitives~\cite{appel-sha}.

One of the main benefits of foundational verification is that it
provides high levels of assurance. The interactive theorem prover
makes sure that proofs of program properties are done in full and
complete detail, without any implicit assumptions or forgotten proof
obligations. Furthermore, once a proof is completed, foundational
proof assistants can generate a representation of the proof in a
foundational logic; these proofs can be checked with a small
kernel. In this setting only the kernel needs to be trusted (as
opposed to the entire proof assistant), leading to a small trusted
computing base. As an example of this high-level of assurance, a study
of compilers~\cite{csmith} has shown that CompCert~\cite{compcert}, a
compiler proved correct in the Coq proof assistant, is significantly
more robust than its non-verified counterparts.

Unfortunately, the benefits of foundational verification come at a
great cost. The process of performing proofs in a proof assistant is
extremely laborious. CompCert~\cite{compcert} took 6 person-years and
100,000 lines of Coq to write and verify, and seL4~\cite{sel4}, which
is a verified version of a 10,000 line operating system, took 22
person-years to verify. The sort of manual effort is one of the main
impediments to the broader adoption of proof assistants.

In this paper, we present \name{}, a novel system that uses machine
learning to help alleviate the manual effort required to complete
proofs in an interactive theorem prover. \name{} trains on existing
proofs to learn models. \name{} then incorporates these learned models
in a tree search process to complete proofs. The source of \name is
publicly available on GitHub~\footnote{https://github.com/UCSD-PL/proverbot9001}.

The main contribution of this paper is bringing domain knowledge
  to the feature engineering, model architecture, and search procedures
  of machine-learning based systems for interactive theorem proving.
In particular, our work distinguishes itself from prior work on machine learning for proofs in three ways:
\begin{enumerate}
  \item A two part tactic-prediction model, in which prediction of tactic arguments is primary and informs prediction of tactics themselves.
  \item An argument prediction architecture which makes use of recurrent neural networks over sequential representations of terms.
  \item Several effective tree pruning techniques inside of a
    prediction-guided proof search.
\end{enumerate}

We tested \name{} end-to-end by training on the proofs from 162 files from CompCert,
  and testing on the proofs from 13 files\footnote{This training/test split comes from splitting the dataset
    90/10, and then removing from the test set files that don't contain proofs.}.
When combined with solver-based tooling (which alone can only solve 7\% of proofs),
\name{} can automatically produce proofs for \PHPercent of the theorem statements
in our test dataset (\PH/501).
In our default configuration without external solvers, \name{} solves
(produces a checkable proof for) 19.36\% (97/501) of the proofs in our
test set, which is a nearly 4X improvement over the previous state of
the art system that attempts the same task~\cite{coqgym}. Our model
is able to reproduce the tactic name from the solution 32\% of the
time; and when the tactic name is correct, our model is able to
predict the solution argument 89\% of the time.
We also show that Proverbot9001 can be trained on one project
  and then effectively predict on another project.

 \section{Background}
\label{sec:background}

\subsection{Foundational Verification}
Program verification is a well studied problem
  in the programming languages community.
Most work in this field falls into one of two categories:
  solver-backed automated (or semi-automated) techniques,
  where a simple proof is checked by a complex procedure;
  and foundational logic based techniques,
  where a complex proof is checked by a simple procedure.

While research into solver-backed techniques
  has produced fully-automated tools in many domains,
  these approaches are generally incomplete,
  failing to prove some desirable propositions.
When these procedures fail,
  it is often difficult or impossible for a user to complete the proof,
  requiring a deep knowledge of the automation.
In contrast,
  foundational verification techniques require a heavy initial proof burden,
  but scale to any proposition without requiring a change in proof technique.
However, the proof burden of foundational techniques can be prohibitive;
  CompCert, a large and well-known foundationally verified compiler,
  took 6 person-years of work to verify~\cite{compcert-experience},
  with other large verification projects sporting similar proof burdens.

\subsection{Interactive Theorem Provers}
Most foundational (and some solver-backed) verification
  is done in an \textit{interactive} theorem prover.
Interactive theorem provers allow the user to define proof goals
  alongside data and program definitions,
  and then prove those goals interactively,
  by entering commands which manipulate the proof context.
The name and nature of these commands varies by the proof assistant,
  but in many foundational assistants,
  these commands are called ``tactics'',
  and coorespond to primitive proof techniques like ``induction'',
  as well as search procedures like ``omega''
  (which searches for proofs over ring-like structures).
Proof obligations in such proof assistants take the form of a set of hypotheses
  (in a Curry-Howard compatible proof theory, bound variables in a context),
  and a goal (a target type);
  proof contexts may consist of multiple proof obligations.

\subsection{Machine Learning and Neural Networks}

  \begin{figure}
    \includegraphics[width=\columnwidth]{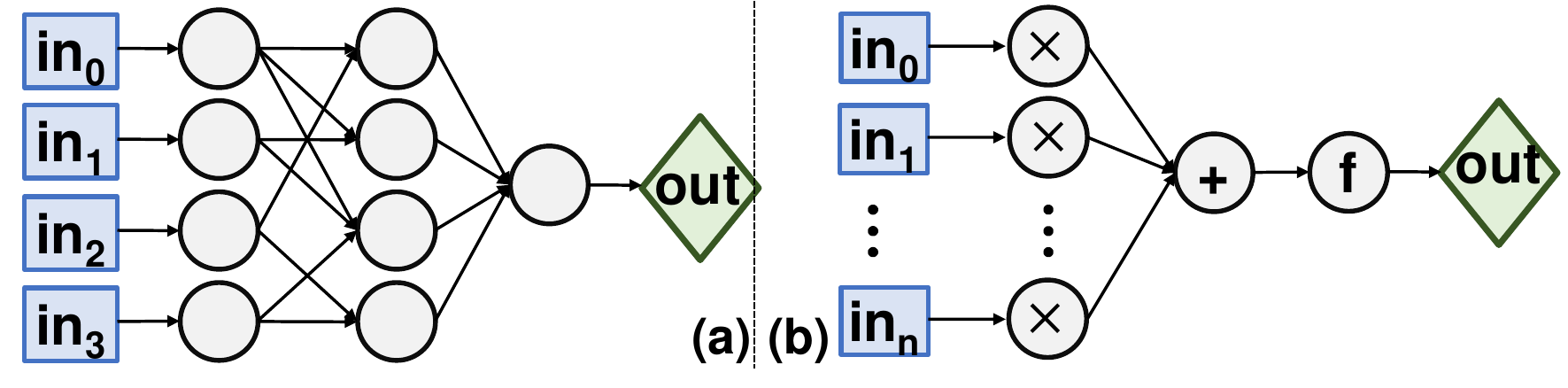}
    \caption{(a) A feed-forward neural network, where each individual gray circle is a
      perceptron (b) An individual perceptron, which multiplies all the inputs by weights, sums up the results, and then applies a non-linear function $f$.}
    \label{fig:ff}
  \end{figure}

  \begin{figure}
    \includegraphics[width=\columnwidth]{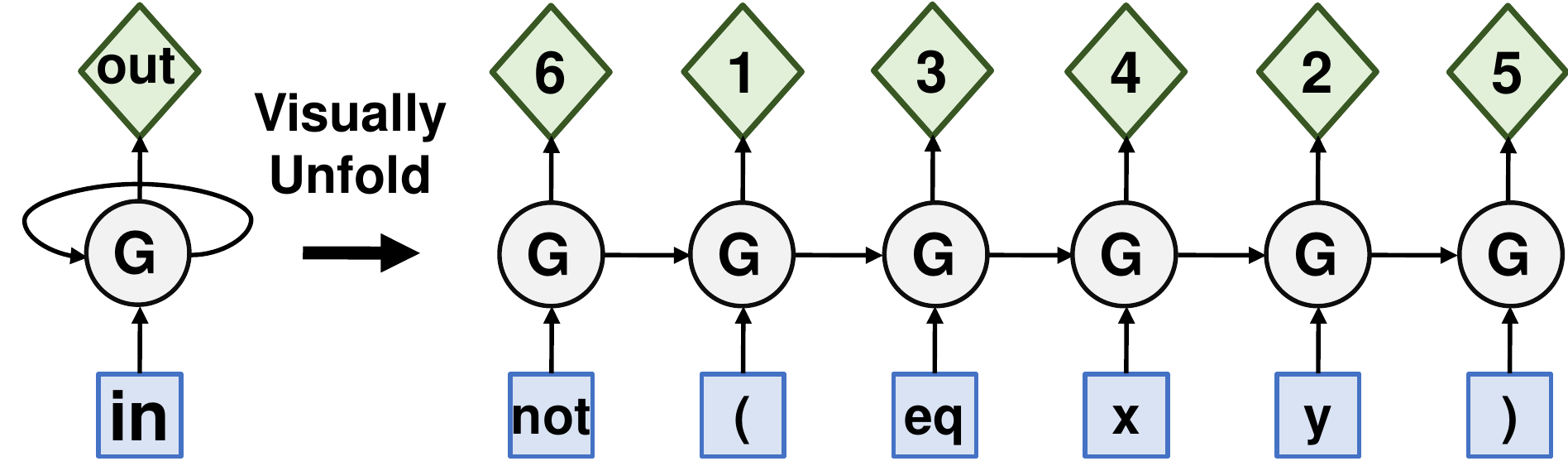}
    \caption{A recurrent neural network. Inputs are in blue boxes at
      the bottom, and each iteration produces an output value, as well
      as a new state value for the next iteration.}
    \label{fig:rnn}
  \end{figure}
Machine learning is an area of computer science dating back to the 1950s.
In problems of supervised learning,
  the goal is to learn a function from labeled examples of input-output pairs.
Models for supervised learning
  parameterize a function from inputs to outputs
  and have a procedure to update the parameters from a data set of labeled examples.
Machine learning has traditionally been applied to problems such as
  handwriting recognition, natural language processing, and recommendation systems.

Neural Networks are a particular class of learned model
  where layers of nodes are connected together
  by a linear combination and a non-linear activation function,
  to form general function approximators.
Neural Networks have a variety of structures,
  some forming a straightforward ``stack'' of nodes
  with some connections removed (convolutional),
  and others, such as those used for natural language processing,
  using more complex structures like loops.

We will make use of two different kinds of neural networks: feed-forward networks and recurrent neural networks.
\Cref{fig:ff}(a) shows the structure of a feed-forward network, where each gray circle is a perceptron, and \Cref{fig:ff}(b) shows individual structure of a perceptron.

\Cref{fig:rnn} shows the structure of a recurrent neural network (RNN). Inputs are shown in blue, outputs in green and computational nodes in gray.
The computational nodes are Gated Recurrent Network nodes,
  GRU for short,
  a commonly used network component with two inputs and two outputs~\cite{gru}.
The network is recurrent because it feeds back into itself,
  with the state output from the previous iteration
  feeding into the state input of the next iteration.
When we display an RNN receiving data, we visually unfold the RNN, as shown on the right side of \Cref{fig:rnn}, even though in practice there is still only one GRU node.
The right side of \Cref{fig:rnn} shows an example RNN that processes tokens of a Coq goal, and produces some output values.
 \section{Overview}
\label{sec:overview}

\begin{figure}
\includegraphics[width=\columnwidth]{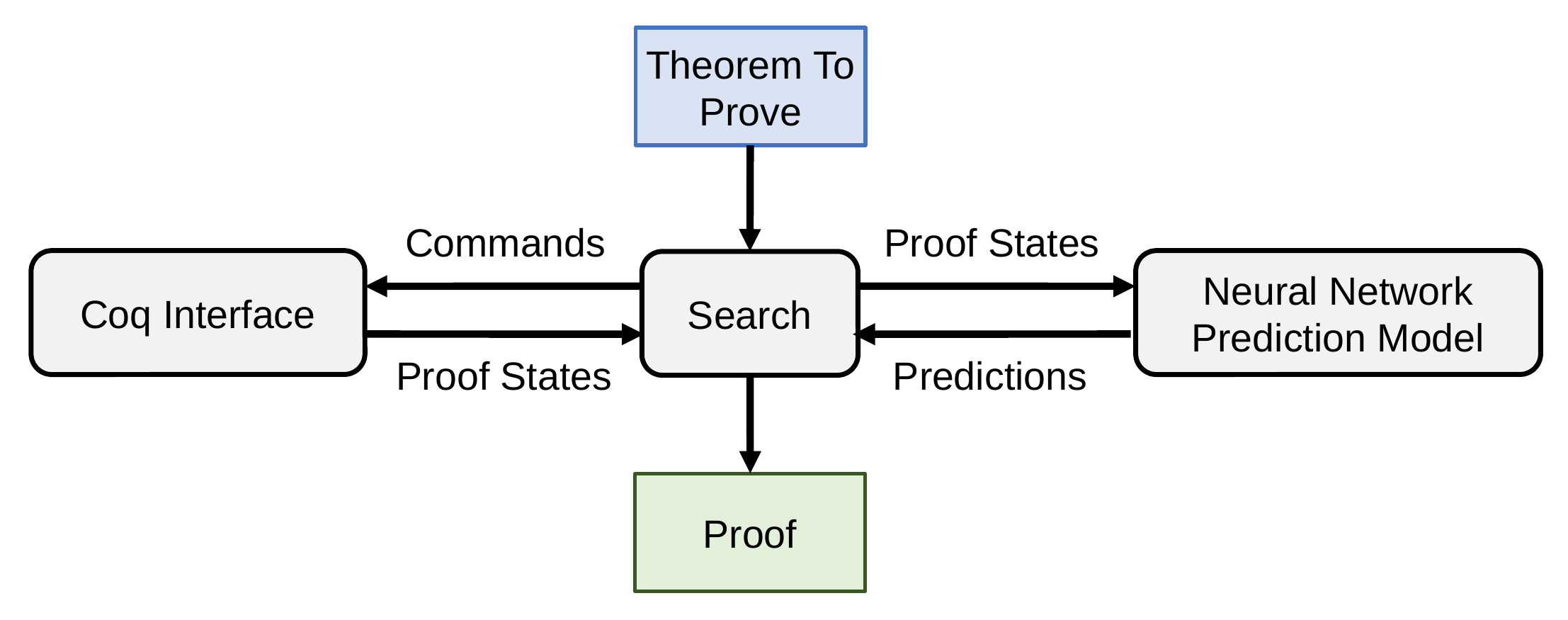}
\caption{The overall architecture of \name, built using CoqSerapi, Python, and PyTroch.}
\label{fig:overview-diagram}
\end{figure}

In this section, we'll present \name's prediction and search process
  with an example from CompCert.
You can see the top-level structure of \name in \Cref{fig:overview-diagram}.

Consider the following theorem from the CompCert compiler:
{
\footnotesize
\begin{minted}[bgcolor=lightgray]{coq}
Definition binary_constructor_sound
    (cstr: expr -> expr -> expr)
    (sem: val -> val -> val) : Prop :=
  forall le a x b y,
  eval_expr ge sp e m le a x ->
  eval_expr ge sp e m le b y ->
  exists v, eval_expr ge sp e m le (cstr a b) v
            /\ Val.lessdef (sem x y) v.

Theorem eval_mulhs:
  binary_constructor_sound mulhs Val.mulhs.
Proof.
...
\end{minted}
}

This theorem states that the \verb-mulhs- expression constructor is
sound with respect to the specification \texttt{Val.mulhs}.

\begin{figure}
  \includegraphics[width=\columnwidth]{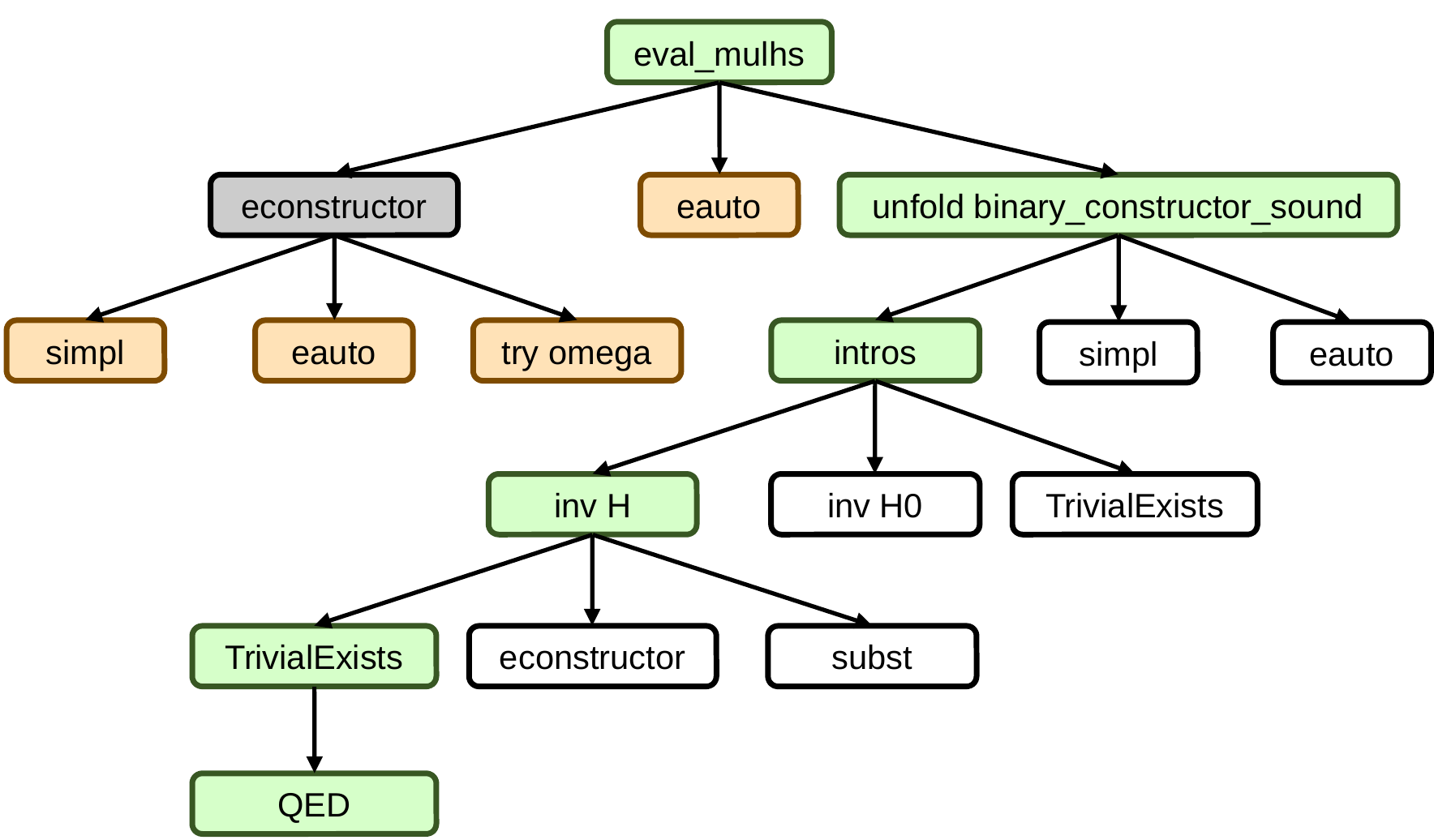}
  \caption{A graph of a \name{} search. In green are the tactics that
    formed part of the discovered solution, as well as the lemma name
    and the QED. In orange are nodes that resulted in a context that
    is at least as hard as one previously found (see \Cref{sec:search}).}
\label{fig:evalmulhs-searchgraph}
\end{figure}

At the beginning of the proof of \texttt{eval_mulhs},
  \name{} predicts three candidate tactics,
  \coqinline{econstructor}, \coqinline{eauto},
  and \coqinline{unfold binary_constructor_sound}.
Once these predictions are made, \name{} tries running all three,
  which results in three new states of the proof assistant.
In each of these three states,
  \name{} again makes predictions for what the most likely tactics are to apply next.
These repeated predictions create a search tree,
  which \name{} explores in a depth first way.
The proof command predictions that \name{} makes are ordered by likelihood,
  and the search explores more likely branches first.

\Cref{fig:evalmulhs-searchgraph} shows the resulting search tree for \texttt{eval_mulhs}.
The nodes in green are the nodes that produce the final proof.
Orange nodes are predictions
  that fail to make progress on the proof (see \Cref{sec:search});
  these nodes are not expanded further.
All the white nodes to the right of the green path are not explored,
  because the proof in the green path is found first.
 \section{Definitions}

In the rest of the paper, we will describe the details of how \name{} works.
We start with a set of definitions that will be used throughout.
In particular, Figure~\ref{fig:defns} shows the formalism we will use to represent the state of an in-progress proof.
\begin{figure}
\begin{tabular}{ll}
$\Tactics$ & Tactics \\
$\Args$ & Tactic arguments\\
$\Commands = \Tactics \times \Args$ & Proof commands \\
$\Idents$ & Identifiers\\
$\Props$ & Propositions \\
$\Goals = \Props$ & Goals \\
$\Hypotheses = \Idents \times \Props$ & Hypotheses \\
$\Obligations = [\Hypotheses] \times \Goals$ & Obligations \\
$\States = [\Obligations \times [\Commands]] $ & Proof states \\
\end{tabular}
\caption{Formalism to model a Proof Assistant}
\label{fig:defns}
\end{figure}
A tactic $\tac \in \Tactics$ is a tactic name.
An argument $\argu \in \Args$ is a tactic argument.
For simplicity of the formalism, we assume that all tactics take zero or one arguments.
We use $\Idents$ for the set of Coq identifiers, and $\Props$ for the set of Coq propositions.
A \emph{proof state} $\st \in \States$ is a state of the proof assistant, which consists of a list of obligations along with their proof command history.
We use $[X]$ to denote the set of lists of elements from $X$.
An obligation is a pair of: (1) a set of hypotheses (2) a goal to prove.
A hypothesis is a proposition named by an identifier, and a goal is a proposition.

\section{Predicting a Single Proof Step}
\label{sec:prediction}

We start by explaining how we predict individual steps in the proof.
Once we have done this, we will explain how we use these proof command predictions to guide a proof search procedure.

We define $\ScoringFuncs[\tac]$ to be a scoring function over $\tac$, where larger scores are preferred over smaller ones:
$$
\ScoringFuncs[\tac] = \tac \rightarrow \Reals
$$
We define a $\tac$-predictor $\Predictors[\tac]$ to be a function that takes a proof state $\st \in \States$ (\emph{i.e.} a state of the proof assistant under which we want to make a prediction) and returns a scoring function over $\tac$. In particular, we have:
$$\Predictors[\tac] = \States \rightarrow \ScoringFuncs[\tac]$$
Our main predictor $\PState$ will be a predictor of the next step in the proof, \emph{i.e.} a predictor for proof commands:
$$
\PState: \Predictors[\Tactics \times \Args]
$$
We divide our main predictor into two predictors, one for tactics, and one for arguments:
$$\PTac: \Predictors[\Tactics]$$
$$\PArg: \Tactics \rightarrow \Predictors[\Args]$$
Our main predictor $\PState$ combines $\PTac$ and $\PArg$ as follows:
$$
\PState(\st) = \lambda (\tac, \argu)~.~\PTac(\st)(\tac)~\Combine~\PArg(\tac)(\st)(\argu)
$$
where $\Combine$ is an operator that combines the scores of the tactic and the argument predictors.
We now describe the three parts of this prediction architecture in turn: $\PTac$, $\PArg$, and $\Combine$.

\subsection{Predicting Tactics ($\PTac$)}
\label{ssec:ptac}

To predict tactics,
  \name{} uses of a set of manually engineered features
  to reflect important aspects of proof prediction:
  (1) the head of the goal as an integer
  (2) the name of the previously run tactic as an integer
  (3) a hypothesis that is heuristically chosen (based on string similarity to goal)
      as being the most relevant to the goal
  (4) the similarity score of this most relevant hypothesis.

These features are embedded into a continuous vector of 128 floats
  using a standard word embedding,
  and then fed into a fully connected feed-forward neural network
  (3 layers, 128 nodes-wide) with a softmax (normalizing) layer at the end,
  to compute a probability distribution over possible tactic names.
This architecture is trained on 153402 samples
  with a stochastic gradient descent optimizer.

\begin{figure}
\includegraphics[width=0.7\columnwidth]{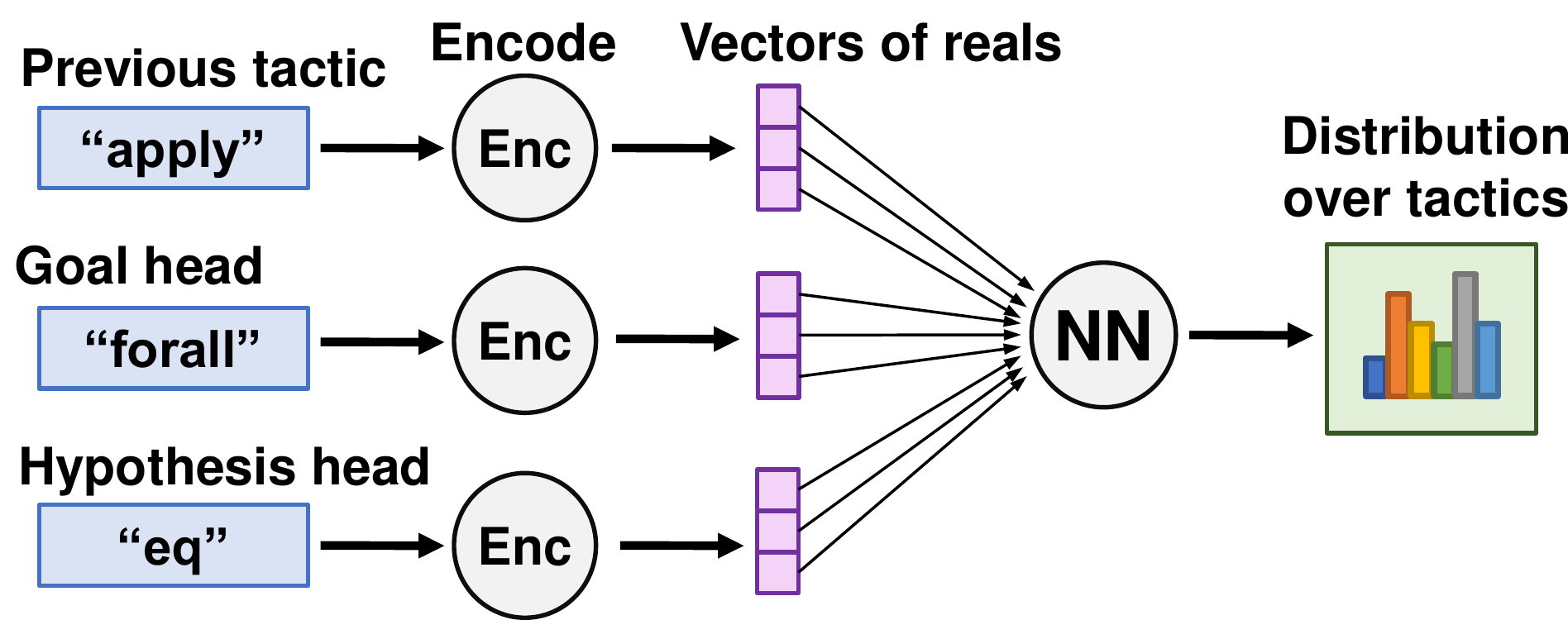}
\caption{\name{}'s model for predicting tactics. Takes as input three
  features for each data point: the previous tactic run, the head
  token of the goal, and of the most relevant hypothesis (see
  \Cref{ssec:ptac}). We restrict the previous tactic feature to the 50
  most common tactics, and head tokens on goal and hypothesis to the
  100 most common head tokens.}
\label{fig:tactic-model}
\end{figure}

The architecture of this model is shown in \Cref{fig:tactic-model}.
Blue boxes represent input;
  purple boxes represent intermediate encoded values;
  green boxes represent outputs;
  and gray circles represent computations.
The NN circle is the feed-forward Neural Network mentioned above.
The Enc circle is a word embedding module.

\subsection{Predicting Tactic Arguments ($\PArg$)}
\label{ssec:arguments}

Once a tactic is predicted, \name{} next predicts arguments.
Recall that the argument predictor is a function $\PArg: \Predictors[\Args]$.
In contrast to previous work, our argument model is a prediction architecture in its own right.

\name{} currently predicts zero or one tactic arguments;
However, since the most often-used multi-argument Coq tactics can be desugared to sequences of single argument tactics (for example ``\coqinline{unfold a, b}'' to ``\coqinline{unfold  a. unfold b.}''), this limitation does not significantly restrict our expressivity in practice.

\name{} makes three kinds of predictions for arguments:
  \emph{goal-token} arguments, \emph{hypothesis} arguments,
  \emph{lemma} arguments:

\emph{Goal-token} arguments are arguments that are a single token in the goal;
  for instance, if the goal is \coqinline{not (eq x y)},
  we might predict \coqinline{unfold not},
  where \coqinline{not} refers to the first token in the goal.
In the case of tactics like \coqinline{unfold} and \coqinline{destruct},
  the argument is often (though not always) a token in the goal.

\emph{Hypothesis} arguments are identifiers referring to a hypothesis in context.
For instance,
if we have a hypothesis \coqinline{H} in context,
with type \coqinline{is_path (cons (pair s d) m)},
we might predict \coqinline{inversion H},
  where \coqinline{H} refers to the hypothesis,
  and \coqinline{inversion} breaks it down.
In the case of tactics like \coqinline{inversion} and \coqinline{destruct},
  the argument is often a hypothesis identifier.

Finally, \emph{lemma} arguments are identifiers referring to a previously defined proof.
These can be basic facts in the standard library, like
\begin{minted}[bgcolor=lightgray]{coq}
plus_n_0 : forall n : nat, n = n + 0
\end{minted}
or a lemma from the current project,
  such as the \coqinline{eval_mulhs} described in the overview.
In \name{}, lemmas are considered from a subset of the possible
  lemma arguments available in the global context,
  in order to make training tractable.
\name{} supports several different modes for determining this subset;
  by default we consider lemmas defined previously in the current file.

\begin{figure}
\includegraphics[width=\columnwidth]{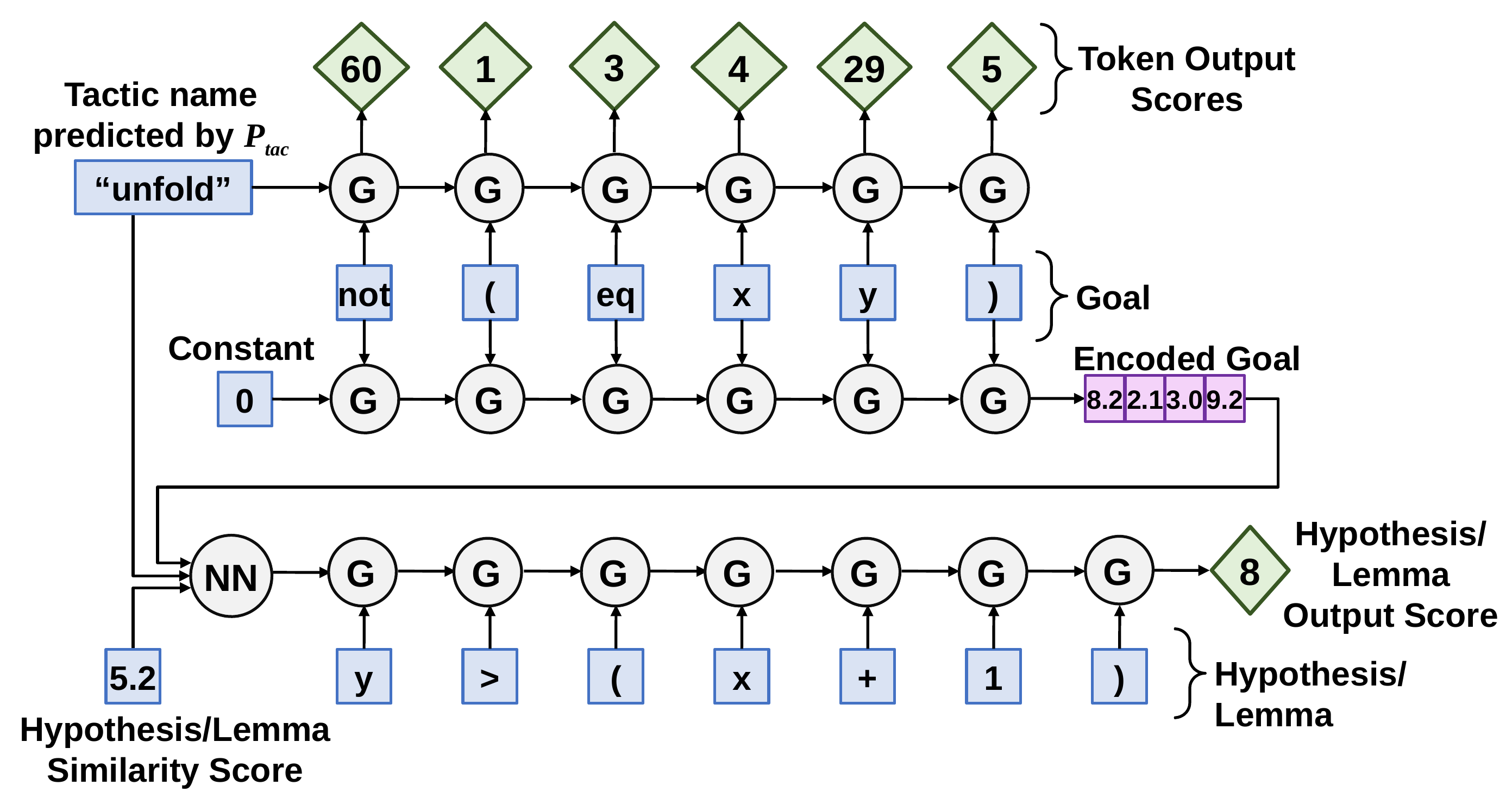}
\caption{The model for scoring possible arguments.}
\label{fig:arguments}
\end{figure}

The architecture of the scoring functions for these argument types is shown in \Cref{fig:arguments}.
One recurrent neural network (RNN) is used to give scores to each hypothesis and lemma
  by processing the type of the term, and outputting a final score.
A different RNN is then used to process the goal, assigning a score to each token in processes.

As before, blue boxes are inputs; purple boxes are encoded values; green diamonds are outputs, in this case scores for each individual possible argument; and gray circles are computational nodes.
The GRU nodes are Gated Recurrent Units~\cite{gru}.
The NN node is a feed-forward neural network.

For illustration purposes, \Cref{fig:arguments} uses an example to provide sample values.
Each token in the goal is an input -- in \Cref{fig:arguments} the goal is ``\coqinline{not (eq x y)}''.
The tactic predicted by $\PTac$ is also an input -- in \Cref{fig:arguments} this tactic is ``\coqinline{unfold}''.
The hypothesis that is heuristically closest to the goal
  (according to our heuristic from \Cref{ssec:ptac}) is also an input,
  one token at a time being fed to a GRU.
In our example,
  let's assume this closest hypothesis is ``\coqinline{y > (x+1)}''.
The similarity score of this most relevant hypothesis is an additional input --
  in \Cref{fig:arguments} this score is $5.2$.

There is an additional RNN (the middle row of GRUs in \Cref{fig:arguments}) which encodes the goal as a vector of reals.
The initial state of this RNN is set to some arbitrary constant, in this case $0$.

The initial state of the hypothesis RNN (the third row of GRUs in \Cref{fig:arguments}) is computed using a feed-forward Neural Network (NN).
This feed-forward Neural Network takes as input the tactic predicted by $\PTac$, the goal encoded as a vector of reals, and the similarity score of the hypothesis.

The architecture in \Cref{fig:arguments} produces one output score for each token in the goal and one output score for the hypothesis.
The highest scoring element will be chosen as the argument to the tactic.
In \Cref{fig:arguments}, the highest scoring element is the ``\coqinline{not}'' token, resulting in the proof command ``\coqinline{unfold not}''.
If the hypothesis score (in our example this score is $8$) would have been the highest score, then the chosen argument would be the identifier of that hypothesis in the Coq context.
For example, if the identifier was \coqinline{IHn} (as is sometimes the case for inductive hypotheses), then the resulting proof command would be ``\coqinline{unfold IHn}''.

\subsection{Combining Tactic and Argument Scores ($\Combine$)}

The $\Combine$ operator attempts to provide a balanced combination
  of tactic and argument prediction,
  taking both into account even across different tactics.
The operator works as follows. We pick the $n$ highest-scoring
tactics and for each tactic the $m$ highest-scoring arguments. We then score
each proof command by multiplying the tactic score and the argument score,
without any normalization. Formally, we can implement this approach by defining
$\Combine$ to be multiplication, and by not normalizing the probabilities
produced by $\PArg$ until all possibilities are considered together.

Because we don't normalize the probabilities of tactics, the potential
arguments for a tactic are used in determining the eligibility of the
tactic itself (as long as that tactic is in the top $n$). This forms
one of the most important contributions of our work: the argument
selection is primary, with the tactic prediction mostly serving to
help prune its search space.

\subsection{Putting It All Together}

\begin{figure*}
\includegraphics[width=\textwidth]{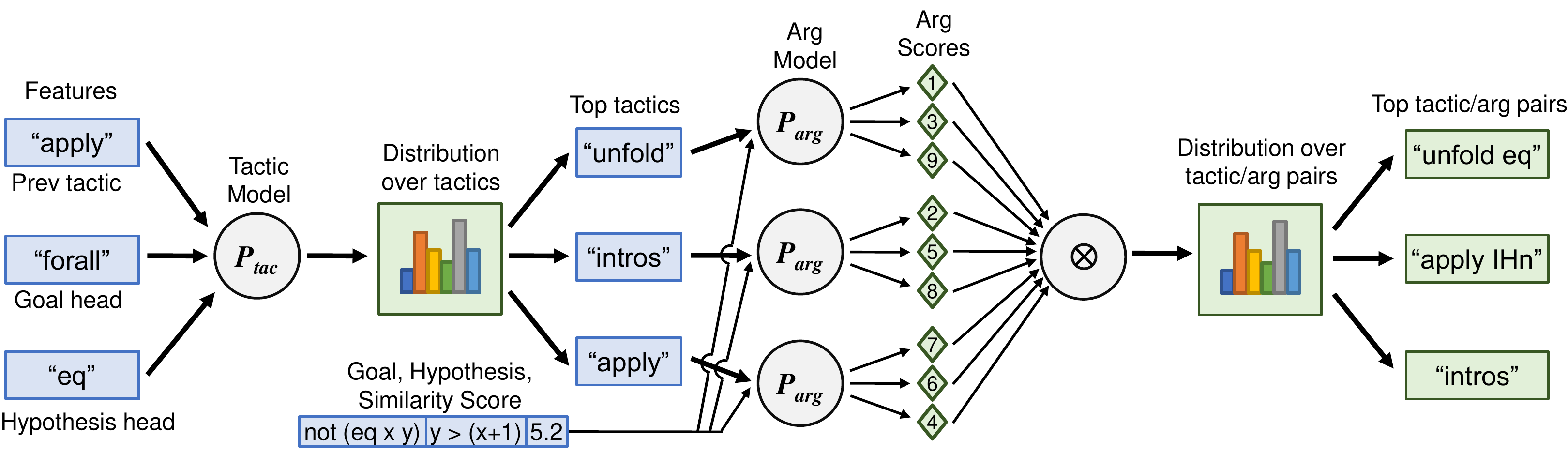}
\caption{The overall prediction model, combining the tactic prediction and argument prediction models.}
\label{fig:overall-model}
\end{figure*}

The overall architecture that we have described is shown in \Cref{fig:overall-model}.
The $\PTac$ predictor (whose detailed structure is shown in \Cref{fig:tactic-model}) computes a distribution over tactic using three features as input: the previous tactic, head
constructor of goal, and head constructor of the hypothesis deemed most relevant.
Then, for each of the top tactic predicted by $\PTac$, the $\PArg$ predictor (whose detailed structure is shown in \Cref{fig:arguments}) is invoked.
In addition to the tactic name, the $\PArg$ predictor takes several additional inputs: the goal, the hypotheses in context, and the similarity between each of those hypotheses and the goal.
The $\PArg$ predictor produces scores for each possible argument (in our case one score for each token in the goal, and one score the single hypothesis).
These scores are combined with $\Combine$ to produce an overall scoring of proof commands.

\section{Training}

\begin{figure}
\includegraphics[width=\columnwidth]{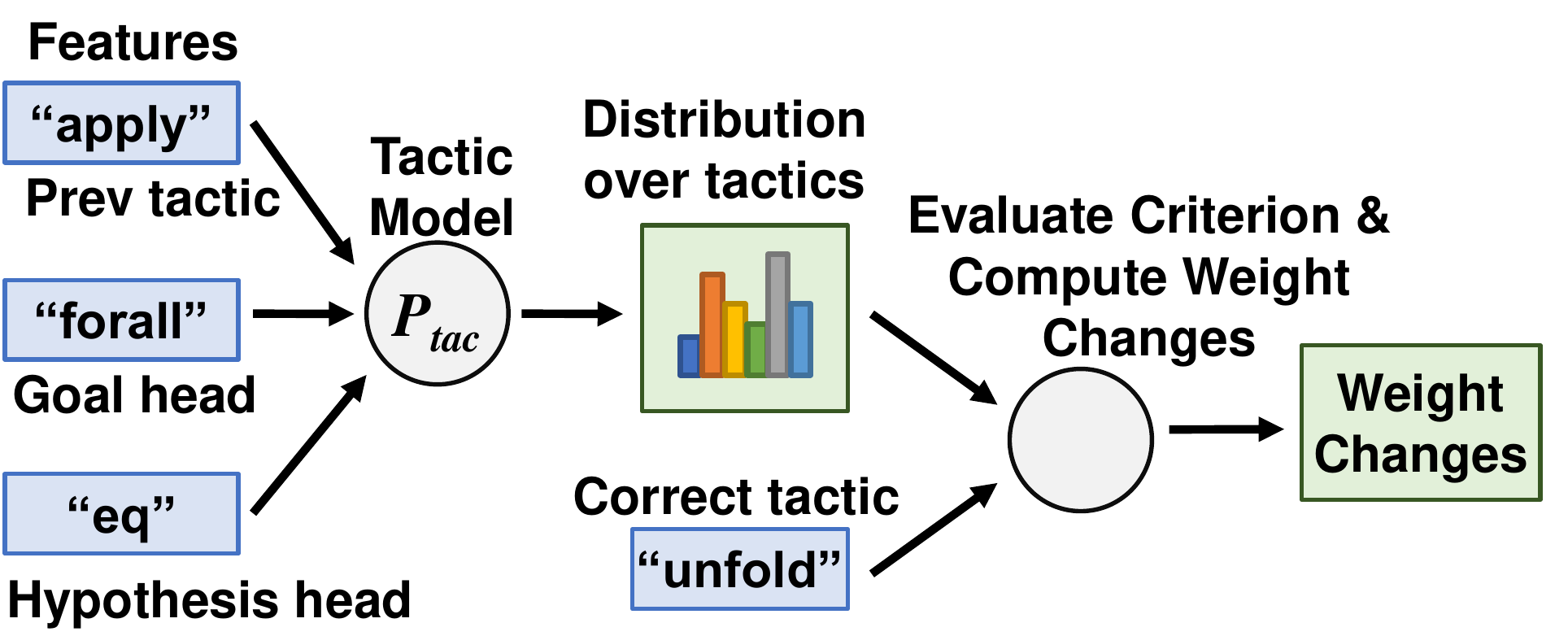}
\caption{The architecture for training the tactic models.}
\label{fig:tactic-training}
\end{figure}
\subsection{Training Architecture}
\Cref{fig:tactic-training} shows the training architecture for the tactic predictor, $\PTac$ (recall that the detailed architecture of $\PTac$ is shown in \Cref{fig:tactic-model}).
The goal of training is to find weights for the neural network that is found inside the gray $\PTac$ circle.
\name{} processes all the Coq theorems in the training set, and steps through the proof of each of these theorems.
\Cref{fig:tactic-training} shows what happens at each step in the proof.
In particular, at each step in the proof, \name{} computes the three features we are training with, and passes these features to the current tactic model to get a distribution over tactics.
This distribution over tactics, along with the correct tactic name (from the actual proof), are passed to a module that computes changes to the weights based on the NLLLoss criterion.
These changes are batched together over several steps of the proof, and then applied to update the tactic model.
Running over all the training data to update the weights is called an epoch, and we run our training over 20 epochs.

\begin{figure}
\includegraphics[width=\columnwidth]{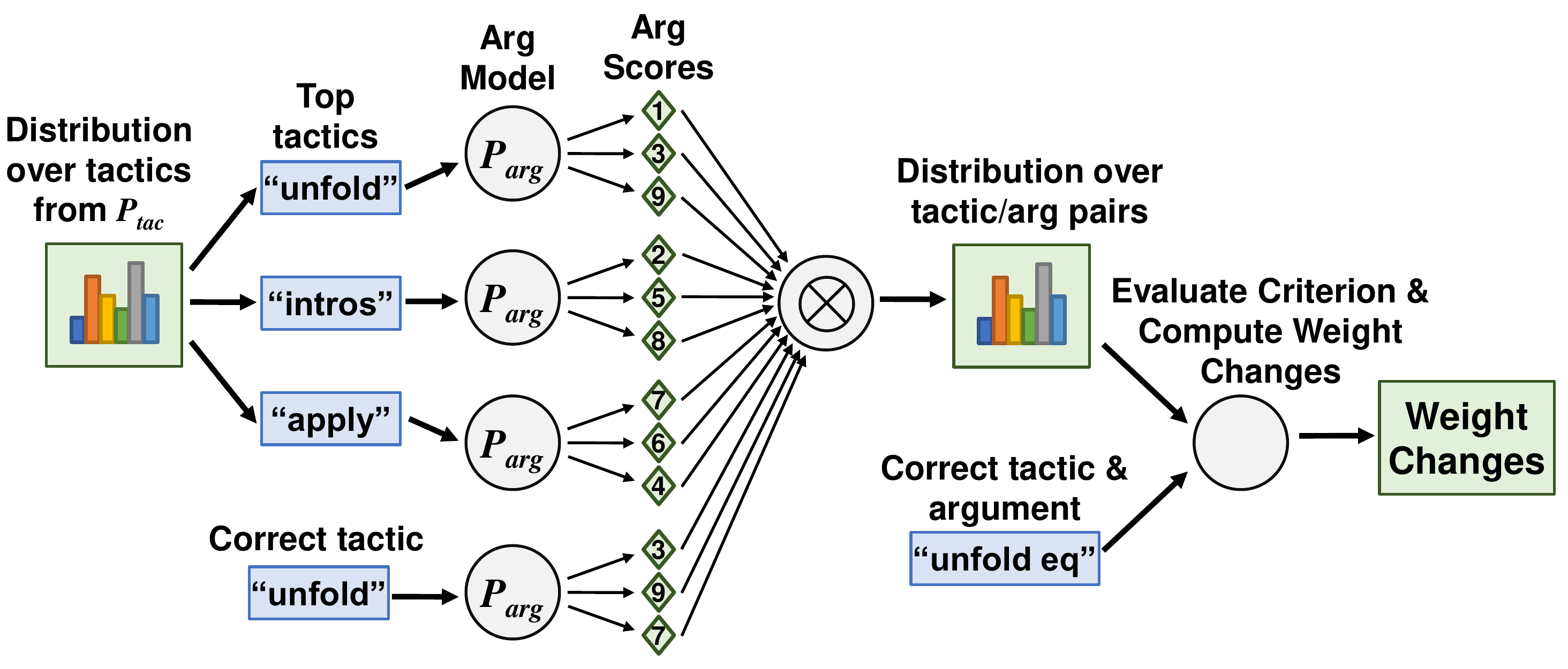}
\caption{The architecture for training the argument
  models. Note that we inject predicted tactics into the input of
  the argument model, instead of just using the correct tactic,
  so that argument scores will be comparable.}
\label{fig:arg-training}
\end{figure}

\Cref{fig:arg-training} shows the training architecture for the argument predictor, $\PArg$ (recall that the detailed architecture of $\PArg$ is shown in \Cref{fig:arguments}).
The goal of training is to find weights for the GRU components in $\PArg$.
Here again, \name{} processes all the Coq theorems in the training set, and steps through the proof of each of these theorems. \Cref{fig:arg-training} shows what happens at each step in the proof.
In particular, at each step in the proof, the current $\PTac$ predictor is run to produce the top predictions for tactic.
These predicted tactic, along with the correct tactic, are passed to the argument model $\PArg$.
To make \Cref{fig:arg-training} more readable, we do not show the additional parameters to $\PArg$ that where displayed in \Cref{fig:overall-model}, but these parameters are in fact also passed to $\PArg$ during training.
Note that it is very important for us to inject the tactics predicted by $\PTac$ into the input of the argument model $\PArg$, instead of using just the correct tactic name.
This allows the scores produced by the argument model to be comparable \emph{across} different predicted tactics.
Once the argument model $\PArg$ computes a score for each possible argument, we combine these predictions using $\Combine$ to get a distribution of scores over tactic/argument pairs.
Finally, this distribution, along with the correct tactic/argument pair is passed to a module that computes changes to the weights based on the NLLLoss criterion.
In our main CompCert benchmark the 153402 tactic samples from the training set are processed for 20 epochs.

\subsection{Learning From Higher-Order Proof Commands}

Proof assistants generally have higher-order proof commands, which are tactics that take other proof commands as arguments;
in Coq, these are called \emph{tacticals}.
One of the most common examples is the (\verb-;-) infix operator which runs the proof command on the right on every sub-goal produced by the tactic on the left.
Another example is the \coqinline{repeat} tactical, which repeats a provided tactic until it fails.

While higher-order proof commands are extremely important for human proof engineers, they are harder to predict automatically because of their generality.
\cut{Indeed, effectively predicting such higher-order tactics would require learning deeper contextual features.
}
While some previous work~\cite{coqgym} attempts to learn directly on data which uses these higher-order proof commands, we instead take the approach of desugaring higher-order proof commands into first-order ones as much as possible;
this makes the data more learnable, without restricting the set of expressible proofs.

For example, instead of trying to learn and predict (\verb|;|) directly, \name{} has a system which attempts to desugar (\verb-;-) into linear sequences of proof commands.
This is not always possible (without using explicit subgoal switching commands), due to propagation of existential variables across proof branches.
\name{} desugars the cases that can be sequenced, and the remaining commands containing (\verb|;|) are filtered out of the training set.

In addition to the (\verb|;|) tactical,
  there are other tacticals in common use in Coq.
Some can be desugared into simpler forms.
For example:
\begin{itemize}
\item ``\coqinline{now <tac>}'' becomes ``\coqinline{<tac>;easy}''.
\item ``\coqinline{rewrite <term> by <tac>}'' becomes ``\coqinline{rewrite <term> ; [ } $\vert$\coqinline{ <tac>]}''
\item ``\coqinline{assert <term> by <tac>}'' becomes ``\coqinline{assert <term> ; [ } $\vert$\coqinline{ <tac>]}''
\end{itemize}

In other cases, like \coqinline{try <tac>} or \coqinline{solve <tac>},
  the tactical changes the behavior of the proof command in a way that cannot be desugared;
  for these we simply treat the prefixed tactic as a separate, learned tactic.
For example, we would treat \coqinline{try eauto} as a new tactic.

\section{Prediction-Guided Search}
\label{sec:search}

Now that we have explained how we predict a single step in the proof, we
describe how \name{} uses these predictions in a proof search.

In general,
proof search works by transitioning the proof assistant into different states by applying proof commands,
and backtracking when a given part of the search space has either been exhausted, or deemed unviable.
Exhaustive proof search in proof assistants is untenable
because the number of possible proof commands to apply is large.
Instead, we use the predictor described above to guide
the search.
Aside from using these predictions, the algorithm is a straightforward depth-limited search, with three subtleties.

\paragraph{First} we stop the search when we find a proof goal
that is at least as hard (by a syntactic definition)
as a goal earlier in the history.
While in general it is hard to formally define
  what makes one proof state harder than another,
  there are some obvious cases which we can detect.
A proof state with a superset of the original obligations will be harder to prove,
and a proof state with the same goal, but fewer assumptions, will be harder to prove.

To formalize this intuition, we define a relation $\ge$ between states
  such that ${\st}_1 \ge {\st}_2$ is meant to capture
  ``Proof state ${\st}_1$ is at least as hard as proof state ${\st}_2$''.
We say that ${\st}_1 \ge {\st}_2$ if and only if for all obligations $O_2$ in ${\st}_2$
  there exists an obligation $O_1$ in ${\st}_1$ such that
  $O_1 {\ge}_o O_2$.
For obligations $O_1$ and $O_2$, we say that
  $O_1 {\ge}_o O_2$ if and only if each hypothesis in $O_1$
  is also a hypothesis in $O_2$,
  and the goals of $O_1$ and $O_2$ are the same.

Since $\ge$ is reflexive,
  this notion allows us to generalize all the cases above
  to a single pruning criteria:
  ``proof command prediction produces a proof state
    which is $\ge$ than a proof state in the history''.

\paragraph{Second} when backtracking, we do not attempt to find a
different proof for an already proven sub-obligation. While in general this can
lead to missed proofs because of existential variables (typed holes filled based on context),
this has not been an issue for the kinds of proofs we have worked with so far.

\paragraph{Third} we had to adapt our notion of search ``depth'' to the
structure of Coq proofs (in which a tactic can produce multiple
sub-obligations).
A na\"ive tree search through the Coq proof space will fail to exploit some of the structure of sub-proofs in Coq.

Consider for example the following two proofs:
\begin{enumerate}
\item \coqinline{intros. simpl. eauto.}
\item \coqinline{induction n. eauto. simpl.}
\end{enumerate}
At first glance, it seems that both of these proofs have a depth of three.
This means that a straightforward tree search (which is blind to the structure of subproofs) would not find either of these proofs if the depth limit were set to two.

However, there is a subtlety in the second proof above which is important (and yet not visible syntactically).
Indeed, the \coqinline{induction n} proof command actually produces two obligations (``sub-goals'' in the Coq terminology).
These correspond to the base case and the inductive case for the induction on \coqinline{n}.
Then \coqinline{eauto} discharges the first obligation (the base case), and \coqinline{simpl} discharges the second obligation (the inductive case).
So in reality, the second proof above really only has a depth of two, not three.

Taking this sub-proof structure into account is important because it allows \name{} to discover more proofs for a fixed depth.
In the example above, if the depth were set to two, and we used a na\"ive search, we would not find either of the proofs.
However, at the same depth of two, a search which takes the sub-proof structure into account would be able to find the second proof (since this second proof would essentially be considered to have a depth of two, not three).

 \section{Evaluation}
\label{sec:evaluation}

This section shows that \name{} is able to successfully solve many proofs.
We also experimentally show that \name{} improves significantly
  on the state-of-the-art presented in previous work.

First, in \Cref{ssec:coqgym-comparison}, we compare experimentally to previous work,
  by running both \name{} and the CoqGym~\cite{coqgym} project on CompCert,
  in several configurations outlined in the CoqGym paper.
Next, in \Cref{ssec:cross},
  we experiment with using the weights learned from one project
  to produce proofs in another.
Then, in \Cref{ssec:length-completion},
  we show the ``hardness'' of proofs that Proverbot9001 is generally able to complete,
  using the length of the original solution as proxy for proof difficulty.
Finally, in \Cref{ssec:individual}, we measure the predictor subsystem, without proof search.
Additional evaluation can be found in the appendix.

Experiments were run on two machines.
Machine A is an Intel i7 machine with 4 cores,
  a NVIDIA Quadro P4000 8BG 256-bit,
  and 20 gigabytes of memory.
Machine B is Intel Xeon E5-2686 v4 machine with 8 cores,
  a Nvidia Tesla v100 16GB 4096-bit,
  and 61 gigabytes of memory.
Experiments were run using GNU Parallel~\cite{gnu-parallel}.

During the development of \name{}, we explored many alternatives,
including n-gram/bag-of-words representations of terms, a variety of features,
and several core models including k-nearest neighbors, support vector machines,
and several neural architectures. While we include here some experiments that
explore high-level design decisions (such as training and testing on the same
projects vs cross project, working with and without solver-based tooling,
modifying the search depth and width, and running with and without
pre-processing), we also note that in the development of a large system tackling
a hard problem, it becomes intractable to evaluate against every possible
permutation of every design decision. In this setting, we are still confident in
having demonstrated a system that works for the specific problem of generating
correctness proof with performance that outperforms the state-of-the-art
techniques by many folds.

\subsection{Summary of Results}

\name{}, run using CoqHammer~\cite{coqhammer} and the default configuration,
  is able to produce proofs for \PHPercent of
  the theorem statements in CompCert.
This represents a \xintDigits:=2;\xintthefloatexpr(\PH/58)\relax X improvement
  over the previous state-of-the-art.
Without any external tooling, \name{} can produce proofs for 19.36\%,
  an almost 4X improvement over previous state-of-the-art prediction-based proofs.
Our core prediction model is able to reproduce the tactic name from the solution
  32\% of the time;
  and when the tactic name is correct,
  our model is able to predict the solution argument 89\% of the time.
We also show that Proverbot9001 can be trained on one project
  and then effectively predict on another project.

\subsection{Experimental Comparison to Previous Work}
\label{ssec:coqgym-comparison}

We tested \name{} end-to-end by training on the proofs from 162 files from CompCert,
  and testing on the proofs from 13 different files.
On our default configuration,
  \name{} solves 19.36\% (97/501) of the proofs in our test set.

In addition to running \name{} on CompCert,
  we ran the CoqGym~\cite{coqgym} tool,
  which represents the state of the art in this area,
  on the same dataset in several configurations.

To account for differences in training dataset,
  we ran CoqGym with their original training schema,
  and also our training schema,
  and reported the best of the two numbers.
CoqGym is intended to be combined with a solver based proof-procedure,
  CoqHammer~\cite{coqhammer}, which is run after every proof command invocation.
While our system was not originally designed this way,
  we compare both systems using CoqHammer,
  as well as both systems without.
We also compared our system to using CoqHammer on the initial goal directly,
  which simultaneously invokes Z3~\cite{z3}, CVC4~\cite{cvc4},
  Vampire~\cite{vampire}, and E Prover~\cite{eprover},
  in addition to attempting to solve the goal using a crush-like tactic~\cite{Chlipala}.

\begin{figure}
\begin{tikzpicture}
  \node[anchor=south west,inner sep=0] at (0,0) {\includegraphics[width=\columnwidth]{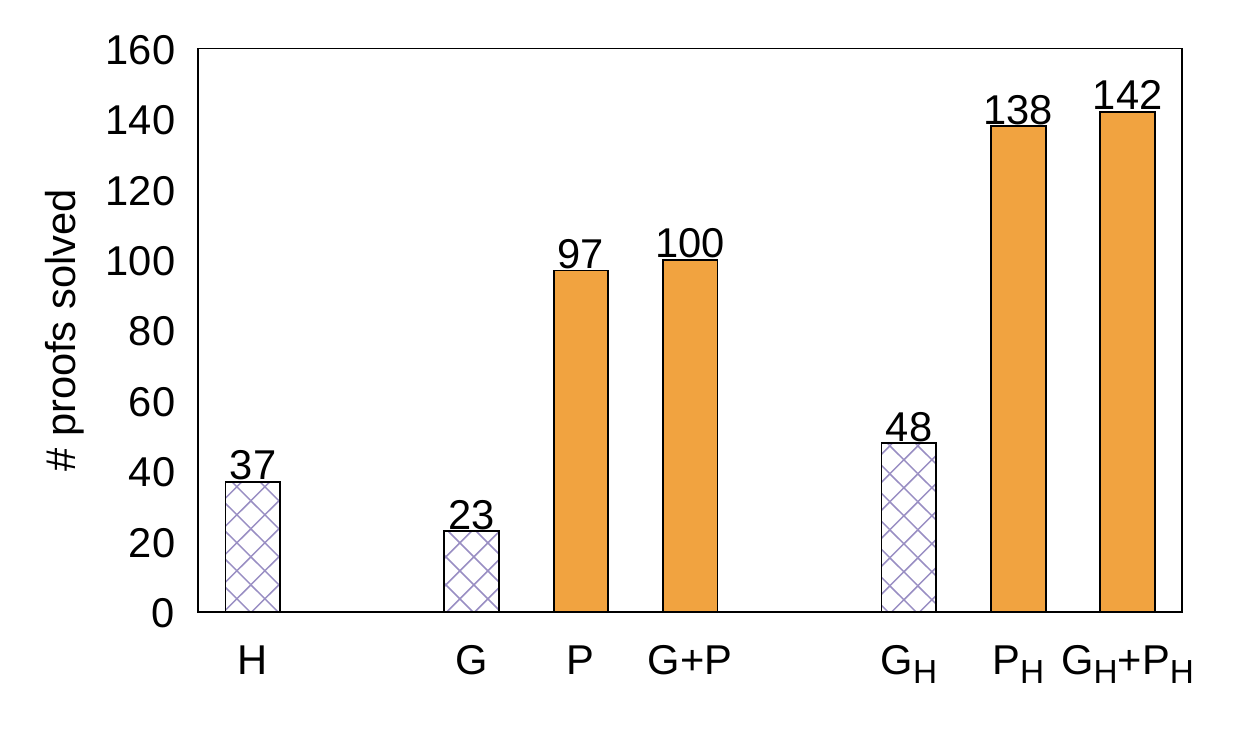}};
  \draw (5.45,4.7) -- (5.45,1);
  \draw (2.45,4.7) -- (2.45,1);
  \node[anchor=south west, inner sep=0] at (3,-1) {\includegraphics[width=\columnwidth/3]{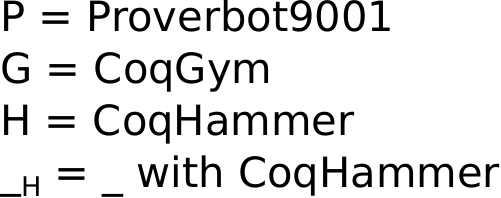}};
\end{tikzpicture}
\caption{A comparison of Proverbot9001 and CoqGym's abilities to
  complete proofs.
H stands for CoqHammer by itself, as a single invocation;
G stands for CoqGym by itself;
P stands for \name{} by itself;
G+P stands for the union of proofs done by G or P;
$\text{G}_\text{H}$ stands for CoqGym with CoqHammer;
$\text{P}_\text{H}$ stands for Proverbot9001 with CoqHammer;
$\text{G}_\text{H}$+$\text{P}_\text{H}$ stands for the union of proofs done by $\text{G}_\text{H}$ or $\text{P}_\text{H}$.}
\label{fig:proofs-solved-bars}
\end{figure}

\Cref{fig:proofs-solved-bars} shows the proofs solved by various configurations.
The configurations are described in the caption.
For all configurations,
  we ran \name{} with a search depth of 6 and a search width of 3
  (see \Cref{sssec:parameters}).
Note that in \Cref{fig:proofs-solved-bars} the bars for H, G, and $\text{G}_\text{H}$ are prior work.
The bars P, G+P and $\text{G}_\text{H}$+$\text{P}_\text{H}$
  are the ones made possible by our work.

When CoqHammer is not used, \name{} can complete nearly 4 times the number of proofs
  that are completed by CoqGym.
In fact, even when CoqGym is augmented with CoqHammer
  \name{} by itself (without CoqHammer) still completes 39 more proofs,
  which is a 67\% improvement (and corresponds to about 8\% of the test set).
When enabling CoqHammer in both CoqGym and \name{}, we see that CoqGym solves 48 proofs whereas \name{} solves \PH{} proofs, which is a \xintDigits:=3;\xintthefloatexpr(\PH/48)\relax X improvement over the state of art.

Finally, CoqGym and \name{} approaches are complementary;
  both can complete proofs which the other cannot.
Therefore, one can combine both tools to produce
  more solutions than either alone.
Combining CoqGym and \name{}, without CoqHammer, allows us to complete 100/501 proofs,
  a proof success rate of 20\%.
Combining \name{} and CoqGym, each with CoqHammer,
  allows us to solve \CHCP{}/501 proofs, a success rate of \number\numexpr((\CHCP * 100)/501)\relax\%.
It's important to realize that,
  whereas the prior state of the art was CoqGym with CoqHammer, at 48 proofs,
  by combining CoqGym and \name{} (both with CoqHammer),
  we can reach a grand total of \CHCP\xspace proofs,
  which is a \xintDigits:=3;\xintthefloatexpr(\CHCP / 48)\relax X improvement
  over the prior state of art.

\subsection{Cross-Project Predictions}
\label{ssec:cross}

To test Proverbot9001's ability to make use of training across projects,
  we used the weights learned from CompCert,
  and ran Proverbot9001 in its default configuration
  on three other Coq projects from the Coq Contrib collection,
  \texttt{concat}, \texttt{float}, and \texttt{zfc}.

\texttt{concat} is a library of constructive category theory proofs,
  which showcases Coq proofs of mathematical concepts instead of program correctness.
The \texttt{concat} library is made of 514 proofs across 105 files;
Proverbot9001 was able to successfully produce a proof for 91 (17.7\%)
  of the extracted theorem statements,
  without the use of CoqHammer.

\texttt{float} is a formalization of floating point numbers, made of 742 proofs across 38 files;
  Proverbot9001 was able to successfully produce a proof for 100 (13.48\%) proofs.

\texttt{zfc} is a formalization of set theory made of 241 proofs across 78 files;
  41 (17.01\%) were successfully completed.

The comparable number for CompCert was 19.36\%.

These results demonstrate
  not only that Proverbot9001 can operate on proof projects in a variety of domains,
  but more importantly that it can effectively transfer training
  from one project to another.
This would allow programmers to use Proverbot9001
  even in the initial development of a project,
  if it had been previously trained on other projects.

\subsection{Original Proof Length vs Completion Rate}
\label{ssec:length-completion}

\begin{figure}
\includegraphics[width=\columnwidth]{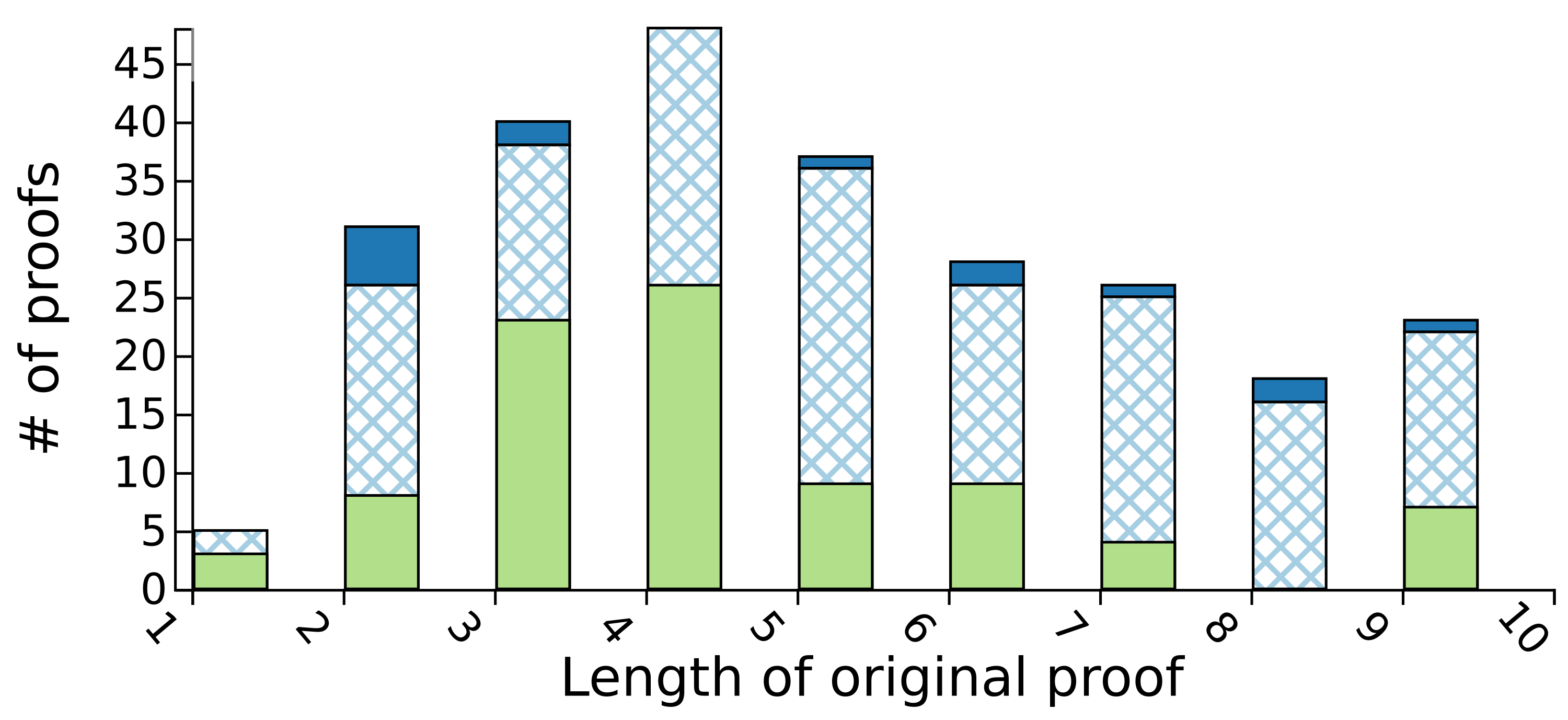}
\caption{A histogram plotting the original proof lengths in proof
  commands vs number of proofs of that length, in three classes, for
  proofs with length 10 or less. From bottom to top: proofs solved,
  proofs unsolved because of depth limit, and proofs where our search
  space was exhausted without finding a solution.}
\label{fig:orig-split-hist-10}
\end{figure}
\begin{figure}
\includegraphics[width=\columnwidth]{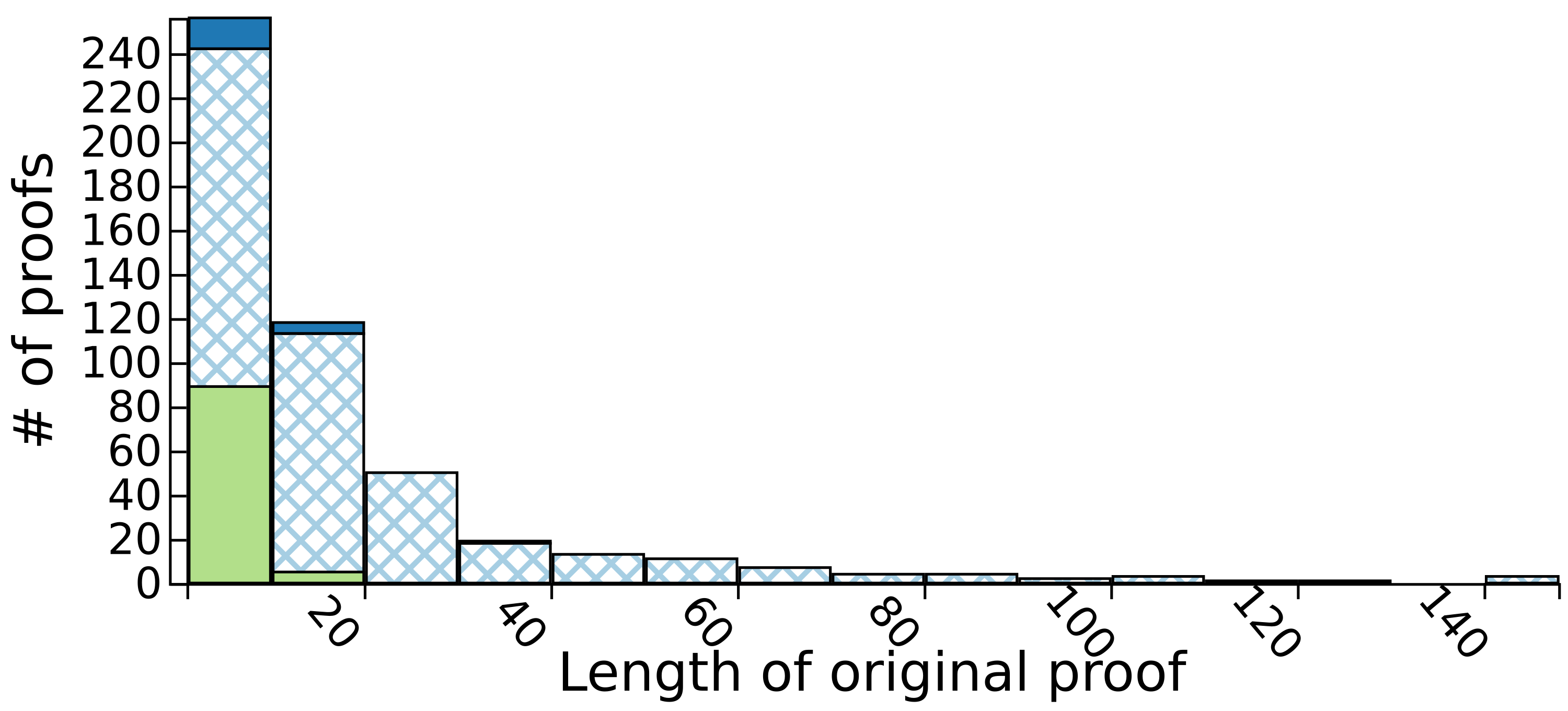}
\caption{A histogram plotting the original proof lengths in proof
  commands vs number of proofs of that length, in three classes. From
  bottom to top: proofs solved, proofs unsolved because of depth
  limit, and proofs where our search space was exhausted without
  finding a solution. Note that most proofs are between 0 and 10 proof
  commands long, with a long tail of much longer proofs.}
\label{fig:orig-split-hist-100}
\end{figure}

In \Cref{fig:orig-split-hist-10} and \Cref{fig:orig-split-hist-100}, we plot a histogram
  of the original proof lengths (in proof commands)
  vs the number of proofs of that length.
We break down the proofs by (from bottom to top)
  number we solve,
  number we cannot solve but still have unexplored nodes,
  and number run out of unexplored nodes before finding a solution.
Note that for the second class (middle bar), it's
  possible that increasing the search depth would allow us to complete
  the proof.
\Cref{fig:orig-split-hist-10} shows proofs of length 10 or below, and \Cref{fig:orig-split-hist-100} shows all proofs, binned in sets of 10.

There are several observations that can be made.
\emph{First}, most original proofs in our test set are less than 20 steps long, with a heavy tail of longer proofs.
\emph{Second}, we do better on shorter proofs.
Indeed, 51\% (256/501) of the original proofs in our test set are ten proof commands or shorter, and of those proofs, we can solve 35\% (89/256), compared to our overall solve rate of 19.36\% (97/501).
\emph{Third}, we are in some cases able to handle proofs whose original length is longer than 10. Indeed, 7 of the proofs we solve (out of 79 solved) had an original length longer than 10.
In fact, the longest proof we solve is originally 25 proof commands long;
  linearized it's 256 proof commands long.
Our solution proof is 267 (linear) proof commands long, comparable to the original proof,
  with frequent case splits.
The depth limit for individual obligations in our search was 6 in all of these runs.
 \section{Related Work}
\label{sec:relatedwork}

\subsection{Program Synthesis}
Program Synthesis is the automatic generation of programs
  from a high-level specification~\cite{dimensions-synthesis}.
This specification can come in many forms,
  the most common being a logical formula over inputs and outputs,
  or a set of input-output examples.
Programs generated can be in a variety of paradigms and languages,
  often domain-specific.
Our tool, \name{}, is a program synthesis tool
  that focuses on synthesis of proof command programs.

Several program synthesis works have used types extensively
  to guide search.
Some work synthesizes programs purely from their types~\cite{complete-completions},
  while other work uses both a type and a set of examples to synthesize programs~\cite{type-and-example, example-and-type}.
In \name{}, the programs being synthesized use a term type as their specification,
  however, the proof command program itself isn't typed using that type,
  rather it must generate a term of that type (through search).

Further work in~\cite{patch-generation} attempts to learn from a set of patches on GitHub,
  general rules for inferring patches to software.
This work does not use traditional machine learning techniques,
  but nevertheless learns from data, albeit in a restricted way.

\subsection{Machine Learning for Code}
Machine learning for modeling code is a well explored area~\cite{survey-ml-for-code},
  as an alternative to more structured methods of modeling code.
Several models have been proposed for learning code,
  such as AST-like trees~\cite{tree-models},
  long-term language models~\cite{deep-code-model},
  and probabilistic grammars~\cite{phog}.
\name{} does not attempt to be so general,
  using a model of programs that is specific to its domain,
  allowing us to capture the unique dependencies of proof command languages.
While the model is simple,
  it is able to model real proofs better than more general models in similar domains
  (see \Cref{ssec:coqgym-comparison}).
Machine learning has been used for various tasks such as
  code and patch generation~\cite{survey-ml-for-code, phog, deep-code-model},
  program classification~\cite{tree-models},
  and learning loop invariants~\cite{learning-invariants}.

\subsection{Machine Learning for Proofs}
While machine learning has previously been explored for various aspects of proof
writing, we believe there are still significant opportunities for improving on
the state-of-the-art, getting closer and closer to making foundational verification
broadly applicable.

More concretely, work on machine learning for
proofs includes: using machine learning to speed up automated solvers~\cite{learning-to-solve},
developing data sets~\cite{holstep, coqgym, holist}, doing
premise selection~\cite{premise1, premise2}, pattern
recognition~\cite{deep-features}, clustering proof data~\cite{ml4pg}, learning
from synthetic data~\cite{gamepad}, interactively suggesting
tactics~\cite{ml4pg, acl2ml}.

Finally, CoqGym attempts to model proofs
  with a fully general proof command and term model expressing arbitrary AST's.
We experimentally compare \name{}'s ability to complete proofs to that of CoqGym
  in detail in \Cref{ssec:coqgym-comparison}
There are also several important conceptual differences.
\emph{First}, the argument model in CoqGym is not as expressive as the one in \name{}.
CoqGym's argument model can predict a hypothesis name, a number between 1 and 4 (which many tactics in Coq interpret as referring to binders, for example \coqinline{induction 2} performs induction on the second quantified variable), or a random (not predicted using machine learning) quantified variable in the goal.
In contrast, the argument model in \name{} can predict any token in the goal, which subsumes the numbers and the quantified variables that CoqGym can predict.
Most importantly because \name{}'s model can predict symbols in the goal, which allows effective unfolding, for example ``\coqinline{unfold eq}''.
\emph{Second}, in contrast to CoqGym, \name{} uses several hand-tuned features for predicting proof commands.
One key example is the previous tactic, which CoqGym does not even encode as part of the context.
\emph{Third}, CoqGym's treatment of higher-order proof commands like ``\coqinline{;}'' is not as effective as \name{}'s.
While neither system can predict ``\coqinline{;}'', \name{} learns from ``\coqinline{;}'' by linearizing them, whereas CoqGym does not.

There is also a recent line of work on doing end-to-end proofs in Isabelle/HOL and
HOL4~\cite{tactictoe,holist,graph-holist}. This work is hard to experimentally
compare to ours, since they use different benchmark sets, proof styles, and
proof languages. Their most recent work~\cite{graph-holist} uses graph
representations of terms, which is a technique that we have not yet used, and
could adapt if proven successful.

Finally, there is also another approach to proof generation, which is to generate
the term directly using language translation models~\cite{proof-translation},
instead of using tactics; however this technique has only been applied to small
proofs due to its direct generation of low-level proof term syntax.
 
\begin{acks}
  We would like to thank Joseph Redmon for his invaluable help building the first \name{} version, Proverbot9000.
\end{acks}

\clearpage
\appendix
\section{Appendix: Additional Evaluation}
\label{sec:appendix}

We now explore more detailed measurements about proof production.

\subsection{Individual Prediction Accuracy}
\label{ssec:individual}

We want to measure the effectiveness of the predictor subsystem that predicts proof command pairs (the $P$ function defined in \Cref{sec:prediction}).
To do this, we broke the test dataset down into individual (linearized) proof commands,
  and ran to just before each proof command to get its prediction context.
Then we fed that context into our predictor,
  and compared the result to the proof command in the original solution.
Of all the proof commands in our test dataset,
  we are able to predict 28.66\% (3784/13203) accurately.
This includes the correct tactic and the correct argument.
If we only test on the proof commands which are in \name{}'s prediction domain,
  we are able to predict 39.25\% (3210/8178) accurately.

During search, our proof command predictor returns the top N tactics for various values of N,
  and all of these proof commands are tried.
Therefore, we also measured how often the proof command in the original proof
  is in the top 3 predictions, and the top 5 predictions.
For all proof commands in the data set, the tactic in the original proof
  is in our top 3 predictions 38.93\% of the time, and in our top 5 predictions 42.66\% of the time.
If we restrict to proof commands in \name{}'s prediction domain, those numbers are 52.17\% and 60.39\%.

\subsection{Argument Accuracy}

Our argument prediction model is crucial to the success of our system,
  and forms one of the main contributions of our work.
To measure its efficacy at improving search is hard,
  because it's impossible to separate its success in progressing a proof
  from the success of the tactic predictor.
However, we can measure how it contributes to individual prediction accuracy.

On our test dataset,
  where we can predict the full proof command in the original proof correctly
  28.66\% of the time,
  we predict the tactic correctly but the argument wrong
  32.24\% of the time.
Put another way,
  when we successfully predict the tactic,
  we can predict the argument successfully with 89\% accuracy.
If we only test on proof commands within \name{}'s prediction domain,
  where we correctly predict the entire proof command 39.25\% of the time,
  we predict the name correctly 41.01\% of the time;
  that is, our argument accuracy is 96\% when we get the tactic right.
It's important to note, however, that many common tactics don't take any arguments,
  and thus predicting their arguments is trivial.

\subsection{Completion Rate in \name{}'s Prediction Domain}

\name{} has a restricted model of proof commands:
  it only captures proof commands with a single argument that is a hypothesis identifier or a token in the goal.
As result,
  it makes sense to consider \name{} within the context of proofs that
  were originally solved with these types of proof commands.
We will call proofs that were originally solved using these types of proof commands \emph{proofs that are in \name{}'s prediction domain}.
There are 79 such proofs in our test dataset (15.77\% of the proofs in the test dataset), and \name{} was able to solve 48 of them.

What is interesting
  is that \name{} is able to solve proofs that are \emph{not} in its prediction domain:
  these are proofs that were originally performed
  with proof commands that are \emph{not} in \name{}'s domain,
  but \name{} found another proof of the theorem
  that \emph{is} in its domain.
This happened for 49 proofs (out of a total of 97 solved proofs).
Sometimes this is because \name{} is able to find a simpler proof command which
  fills the exact role of a more complex one in the original proof;
  for instance, \coqinline{destruct (find_symbol ge id)} in an original proof
  is replaced by \coqinline{destruct find_symbol} in \name{}'s solution.
Other times it is because \name{} finds a proof
  which takes an entirely different path than the original. In fact,
  31 of \name{}'s 97 found solutions are shorter than the original.
It's useful to note that while previous work had a more expressive proof command model,
  in practice it was unable to solve as many proofs
  as \name{} could in our more restricted model.

Together, these numbers indicate that the restricted tactic model used by \name{}
  does not inhibit its ability to solve proofs in practice,
  even when the original proof solution used tactics outside of that model.

\subsection{Data Transformation}

Crucial to \name{}'s performance is its ability to learn from data
  which is not initially in its proof command model,
  but can be transformed into data which is.
This includes desugaring tacticals like \coqinline{now},
  splitting up multi-argument tactics like \coqinline{unfold a, b}
  into single argument ones,
  and rearranging proofs with semicolons into linear
  series of proof commands.
To evaluate how much this data transformation
  contributes to the overall performance of \name{},
  we disabled it, and instead filtered the proof commands
  in the dataset which did not fit into our proof command model.

With data transformation disabled,
  and the default search width (5) and depth (6),
  the proof completion accuracy of \name{} is
  15.57\% (78/501 proofs).
Recall that with data transformation enabled as usual,
  this accuracy is 19.36\%.
This shows that the end-to-end performance of \name{}
  benefits greatly from the transformation of input data,
  although it still outperforms prior work (CoqGym) without it.

When we measure the individual prediction accuracy of our model,
  trained without data transformation,
  we see that its performance significantly decreases
  (16.32\% instead of 26.77\%),
  demonstrating that the extra data produced by preprocessing
  is crucial to training a good tactic predictor.

\subsection{Search Widths and Depths}
\label{sssec:parameters}

Our search procedure has two main parameters,
  a \textit{search width}, and a \textit{search depth}.
The \textit{search width} is how many predictions are explored
  at each context.
The \textit{search depth} is the longest path from the root
  a single proof obligation state can have.

To explore the space of possible depths and widths,
  we varied the depth and width,
  on our default configuration without external tooling.
With a search width of 1 (no search, just running the first prediction),
  and a depth of 6,
  we can solve 5.59\% (28/501) of proofs in our test dataset.
With a search width of 2, and a depth of 6,
  we're able to solve 16.17\% (81/501) of proofs,
  as opposed to a width of 3 and depth of 6, where we can solve 19.36\% of proofs.

To explore variations in depth, we set the width at 3, and varied depth.
With a depth of 2,
  we were able to solve 5.19\% (26/501) of the proofs in our test set.
By increasing the depth to 4,
  we were able to solve 13.97\% (70/501) of the proofs in our test set.
At a depth of 6 (our default),
  that amount goes up to 19.36\% (97/501).

\end{document}